\documentclass[twocolumn]{revtex4}
\usepackage{mathrsfs}
\usepackage{amssymb}
\usepackage[tbtags]{amsmath}
\usepackage{graphicx}
\usepackage[font=small,format=plain, justification=RaggedRight, labelsep=period]{caption}
\usepackage{epsfig,graphicx,times}
\usepackage{subfig}
\usepackage{color}

\begin{document}

\title{Quantum state engineering with flux-biased
Josephson phase qubits by rapid adiabatic passages}
\author{W. Nie, J. S. Huang, X. Shi and L. F. Wei\footnote{
weilianfu@gmail.com, lfwei@swjtu.cn}}
\affiliation{Quantum
Optelectronics Laboratory, Southwest Jiaotong University, Chengdu
610031, China}
\date{\today}

\begin{abstract}
In this paper, the scheme of quantum computing based on
Stark-chirped rapid adiabatic passage (SCRAP) technique [Phys. Rev.
Lett. \textbf{100}, 113601 (2008)] is extensively applied to
implement the quantum-state manipulations in the flux-biased
Josephson phase qubits.
The broken-parity symmetries of bound states in flux-biased
Josephson junctions are utilized to conveniently generate the
desirable Stark-shifts. Then, assisted by various transition pulses,
universal quantum logic gates as well as arbitrary quantum-state
preparations could be implemented.
Compared with the usual $\pi$-pulses operations widely used in the
experiments, the adiabatic population passages proposed here are
insensitive to the details of the applied pulses and thus the
desirable population transfers could be satisfyingly implemented.
The experimental feasibility of the proposal is also discussed.

PACS number(s): 85.25.Cp, 03.67.Lx, 42.50.Dv.
\end{abstract}

\maketitle

\section{Introduction}

Quantum state engineering (QSE) based on superconducting Josephson
circuits (SJCs)~\cite{makhlin} has been stimulated by the
encouraging prospects of quantum computing (QC) and quantum
information processing
(QIP)~\cite{{you},{clarke},{wendin},{devoret}}. The SJCs include the
charge-~\cite{nakamura}, flux-~\cite{{friedman},{mooij}}, and phase
qubits~\cite{{Martinis},{Simmonds},{berkley}} as well as their
variants~\cite{{vion},{koch},{manucharyan}}. Moreover, QSE with SJCs
is also a crucial approach to investigate the fundamental quantum
phenomena, such as geometric phase~\cite{leek} and
non-locality~\cite{{Wei3},{ansmann}}. Note that most of the current
experiments for QSE employ the technique that is sensitive to the
exact design of the applied pulses. A typical example is that the
so-called $\pi$-pulse is usually exactly designed to transfer the
population from one quantum state to another. However, these
duration-sensitive operations might not be practically the most
optimal approaches to implement the desirable quantum manipulations
with sufficiently-high fidelity and efficiency.

Besides the exactly-designed pulse operations, adiabatic passage
(AP) technique developed in atomic physics is also an efficient
strategy for quantum-state control. This technique possesses certain
significant advantages, such as high transfer efficiency, robustness
to environment noises, and less limits on the designs of the
operational pulses.
It is well-known that the main APs include such as the stimulated
Raman AP (STIRAP)~\cite{bergmann}, Stark chirped rapid AP
(SCRAP)~\cite{{yatsenko1},{rangelov},{oberst}}, and piecewise AP
(PAP)~\cite{shapiro}, etc.. Certainly, all of these methods are
competent for various population transfers. But, certain limits
still exist in these APs. For example, STIRAPs are not immune to the
induced ac Stark shifts~\cite{rickes}. Also, PAP demonstrated in
recent experiments still needs a series of ultrafast pulses to
simultaneously control the switches of two pulses and phase
holdings. This further requires the superb operating techniques.
In contrast, SCRAP realized in the recent experiment~\cite{oberst}
only needs relevant controls on the amplitudes of pulses and thus
could be immune to the inhomogeneous level broadening.

AP technique has also been used in SJCs for various applications,
(see, e.g.,~\cite{kis,siewert,xia,song,wei2008}). For example, the
STIRAP technique was utilized to prepare Fock state~\cite{siewert}
of a nanomechanical oscillator coupled to a Cooper-pair box,
two-qubit and three-qubit entangled states in coupled Josephson flux
circuits were proposed~\cite{xia,song}, etc.. In particular, in
Ref.~\cite{wei2008} we proposed an effective approach, by using the
SCRAP technique, to implement the fundamental logic gates without
precisely designing the durations of the operations.
In this paper, we further generalize such an idea to implement the
QSE with flux-biased Josephson circuits, including the single-qubit
phase shift and also the $\sigma_x$-rotation operation, as well as
the two-qubit iSWAP gate.
Our basic idea is to introduce a controllable flux-perturbation to
chirp the transition frequency of a selected flux-biased qubit.
Then, by properly controlling the amplitude ratio of perturbation to
the transition driving, desirable population transfers between the
selected quantum states could be achieved with sufficiently-high
efficiency. Based on these operations, various single- and two-qubit
logic gates in these SJCs could be implemented without exactly
designing the durations of the applied adiabatic pulses.
Furthermore, arbitrary superposition of the logic states could also
been implemented, in principle. Given the flux-biased Josephson
phase qubits and their manipulations have already been demonstrated,
the present proposal should be experimentally feasible.

The paper is organized as follows: In Sec.~II we give a brief review
of the usual $\pi$-pulse coherent excitation, and the SCRAP
technique used in atomic physics is also included. Subsequently, we
give a simple description of the flux-biased phase qubit, a detailed
approach for population adiabatic transfers, as well as controllable
construction of superposition state. In Sec.~III, we discuss how to
implement the two-qubit gate with capacitively-coupled flux-biased
phase qubits by the proposed SCRAP technique. Finally, conclusions
and discussions are given in Sec.~IV.

\section{Population transfers between driven two levels}
Two-level system is a fundamental model in quantum physics. In QIP
the basic information element (i.e., qubit) is encoded by a
well-defined two-level system. There are many approaches to
implement the population transfers between the two levels of the
qubit. Roughly, these approaches can be classified into the
nonadiabatic- and adiabatic passages~\cite{bergmann}.

\subsection{Nonadiabatic population passages: $\pi$-pulse coherent excitations}
The Hamiltonian of a single qubit driven by an external field can be
generally written as ($\hbar\equiv1$)
\begin{equation}
H_{s}(t)=\frac{\omega_{0}}{2}\sigma_{z}+R(t)\sigma_{x},
\label{originalH}
\end{equation}
where $\sigma_{z}$ and $\sigma_{x}$ refer to the Pauli spin
operators and $\omega_{0}$ the transition frequency of the qubit.
$R(t)=\Omega(t)\cos(\upsilon t)$ is the driving term with $\upsilon$
being the pulse frequency and $\Omega(t)=\varepsilon(t)\mu/\hbar$
the Rabi frequency. Here, $\varepsilon(t)$ is the amplitude of the
applied pump pulse, and $\mu$ the matrix element of electric dipole
moment. Suppose that the applied pump $I_{ac}(t)=\varepsilon(t)
\cos(\upsilon t)$ is resonant with the qubit, i.e., $\upsilon =
\omega_{0}$. In the interaction picture, Eq.~(\ref{originalH})
reduces to
\begin{eqnarray}
H_{int}(t)&=&\exp{(it\omega_{0}\sigma_{z}/2)}R(t)\sigma_{x}\exp{(-it\omega_{0}\sigma_{z}/2)}\nonumber\\
&=&\frac{\Omega(t)}{2}\sigma_{x},
\end{eqnarray}
under the usual rotating-wave approximation (RWA). Correspondingly,
the solution to the Schr\"odinger equation
\begin{equation}
i\frac{\partial U_{int}(t)}{\partial t}=H_{int}(t)U_{int}(t),
\end{equation}
can be expressed as
\begin{eqnarray}
U_{int}(t)&=&\exp{\left(-i\int_0^tH_{int}(t')dt'\right)}\nonumber \\
&=&\cos \frac{A(t)}{2}\textbf{I}-i\sin \frac{A(t)}{2}\sigma_{x}.
\end{eqnarray}
Here, $A(t)=\int_0^t \Omega(t')dt'$ and $\textbf{I}$ is an unit
matrix.
This implies that, if the qubit is initially prepared in the ground
state, then at the time $t$ the probability for the qubit evolving
to the excited state is $P_{e}(t)=(1-\cos A(t))/2$.
Therefore, in order to realize complete population inversion, the
pulse area must be precisely designed as $\pi$. Any deviation from
the precise pulse area may result in dynamical error for the
desirable population inversion.

\subsection{Adiabatic population passages: Stark Chirps}

For loosening the above rigorous requirement on exactly designing
pulse area and improving the operational reliability, we add a
controllable perturbation to the Hamiltonian (\ref{originalH}),
i.e.,
\begin{equation}
H'_{s}(t)=\frac{\omega _{0}}{2}\sigma _{z}+R(t)\sigma
_{x}-\frac{\Delta (t)}{2}\sigma _{z}.
\end{equation}
In the interaction picture, the above Hamiltonian reduces to
\begin{equation}
H'_{int}(t)=\frac{1}{2}\left(
\begin{array}{cc}
0 & \Omega (t) \\
\Omega (t) & 2\Delta (t)%
\end{array}%
\right). \label{Hintp}
\end{equation}
Note that this Hamiltonian is as the same as that in the original
SCRAP scheme~\cite{yatsenko1}, wherein the Hamiltonian is derived in
the Schr\"odinger picture. Also, the pump pulse $I_{ac}(t)$ applied
in the present scheme is required to be resonant with the qubit.
Certainly, compared with the transition frequency $\omega_0$ of the
qubit, the controllable Stark-shift term $\Delta(t)$ should be
sufficiently small.
For convenience, we rewrite the Hamiltonian (\ref{Hintp}) as
\begin{eqnarray}
H''_{int}(t)&=&\frac{\epsilon(t)}{2}\left(
\begin{array}{cc}
0 &\frac{ \Omega (t)}{\sqrt{\Delta^{2}(t)+\Omega^{2}(t)}} \\
\frac{\Omega (t)}{\sqrt{\Delta^{2}(t)+\Omega^{2}(t)}} & \frac{2\Delta (t)}{\sqrt{\Delta^{2}(t)+\Omega^{2}(t)}}%
\end{array}%
\right)\nonumber\\
&=&\frac{\epsilon(t)}{2} \left(
\begin{array}{cc}
0 & \sin 2\vartheta(t) \\
\sin 2\vartheta(t) & 2\cos 2\vartheta(t) \\
\end{array}
\right),\label{Hintpp}
\end{eqnarray}
where $\epsilon(t)=\sqrt{\Delta^{2}(t)+\Omega^{2}(t)}$, and
$\tan2\vartheta(t)=\Omega(t)/\Delta(t)$.
The instantaneous eigenvalues of the above time-dependent
Hamiltonian can be straightforwardly written as: $\mu _{\pm
}(t)=(\Delta (t)\pm \sqrt{\Delta ^{2}(t)+\Omega ^{2}(t)})/2$ with
the corresponding eigenvectors,
\begin{subequations}
\begin{equation}
|\lambda _{+}(t)\rangle =\sin \vartheta(t)\,|0\rangle +\cos
\vartheta(t)\,|1\rangle ,
\end{equation}
\begin{equation}
|\lambda _{-}(t)\rangle =\cos \vartheta(t)\,|0\rangle -\sin
\vartheta(t)\,|1\rangle.
\end{equation}
\end{subequations}%
In the new Hilbert space spanned by the vectors $|\lambda
_{-}(t)\rangle$ and $|\lambda _{+}(t)\rangle$, the Hamiltonian
(\ref{Hintp}) reads (see Appendix)
\begin{equation}
H_{new}(t)=\left(
            \begin{array}{cc}
              \mu_{-}(t) & -i\dot{\vartheta}(t)\\
              i\dot{\vartheta}(t) & \mu_{+}(t) \\
            \end{array}
          \right).\label{newH}
\end{equation}
Under the adiabatic approximation,
\begin{equation}
\frac{1}{2}\left|\Omega (t)\frac{d\Delta (t)}{dt}-\Delta (t)\frac{d\Omega (t)%
}{dt}\right|\ll \left( \Delta ^{2}(t)+\Omega ^{2}(t)\right)^{3/2},
\end{equation}%
Hamiltonian Eq. (\ref{newH}) can be further simplified to
\begin{equation}
H_{ad}(t)=\left(
            \begin{array}{cc}
              \mu_{-}(t) & 0 \\
              0 & \mu_{+}(t) \\
            \end{array}
          \right).
\end{equation}
The vanished nondiagonal elements denote that there is not any
transition between the two instantaneous eigenstates
$|\lambda_{-}(t)\rangle$ and $|\lambda_{+}(t)\rangle$.
This implies that the qubit would passage individually along one of
the two adiabatic paths, as long as $\vartheta(t)$ changes slowly.
As a consequence, the generic solution of the system takes the form
\begin{eqnarray}
|\Psi(t)\rangle &= v_{-}(0)\exp(-i\int_{0}^{t}\mu_{-}(t')dt')
|\lambda_{-}(t)\rangle+&\nonumber\\
&\,\,\,v_{+}(0)\exp(-i\int_{0}^{t}\mu_{+}(t')dt')|\lambda_{+}(t)\rangle.&
\end{eqnarray}
Obviously, although the population of an adiabatic state is
conservational for no coupling between adiabatic states, the
components in the states $|0\rangle$ and $|1\rangle$ can still vary
with the time-dependent $\vartheta(t)$. In principle, one can
realize arbitrary population distributions in $|0\rangle$ and
$|1\rangle$ along one selected adiabatic path. As an obvious
advantage, the population transfer presented here is not sensitive
to the pulse area.

When the pump pulse is absent, the Hamiltonian Eq.~(\ref{Hintp})
becomes $H_{z}=\Delta (t)|1\rangle \langle 1|$. This indicates that
a Stark-chirping pulse is sufficient to produce a phase shift gate
$U_{z}(\alpha )=\exp (i\alpha |1\rangle \langle 1|)$ with $\alpha
=-\int_{t_{0}}^{t_{f}}\Delta (t')dt'$:
\begin{eqnarray}
|0\rangle \longrightarrow |0\rangle, \,\, |1\rangle \longrightarrow
e^{i\alpha }|1\rangle.
\end{eqnarray}
This is similar to the idea by lowering the potential to implement
the fast qubit's readout~\cite{cooper}.
This is because that the Stark pulse does not destruct the
population distributions in the states $|0\rangle$ and $|1\rangle$,
but just leads to the phase accumulations~\cite{bialczak}.
By combining the Rabi pulse for transferring the populations between
the two levels and the Stark pulse for phase shift operation, one
can implement arbitrary superposition of the states $|0\rangle$ and
$|1\rangle$, with controllable probabilities and relative phase.

(i) Implement the $\sigma_x$-rotation operation. We design a pulse
sequence shown in Fig.~\ref{singlequbit}(a): apply only the Stark
pulse at the first for satisfying the initial condition
$\vartheta(t_{0})=0$; and then a pump pulse is applied but it
switches off prior to Stark pulse for satisfying the condition
$\vartheta(t_{f})=\pi/2$. This pulse sequence yields the following
population inversion
\begin{eqnarray}
&&\!\!\!\!\!\!\!\!\!\!\!\!\!\!\!\!\!\!\!\!\!\! H_{inv}:\left\{
\begin{split}
|\Psi(t_{0})\rangle=|0\rangle\xrightarrow{|\lambda_{-}(t)\rangle}
|\Psi(t_{f})\rangle=-|1\rangle\\
|\Psi(t_{0})\rangle=|1\rangle\xrightarrow{|\lambda_{+}(t)\rangle}
|\Psi(t_{f})\rangle=\;\;\;|0\rangle
\end{split}
\right.
\end{eqnarray}
along the adiabatic paths $|\lambda_+(t)\rangle$ and
$|\lambda_-(t)\rangle$, respectively. If the qubit resides in
$|0\rangle$ (or $|1\rangle$) originally, it would evolve to the
state $|1\rangle$ (or $|0\rangle$) along the adiabatic path
$|\lambda_{-}(t)\rangle$ (or $|\lambda_{+}(t)\rangle$). After
eliminating the additional phase via a phase-shift gate operation
$H_{z}(\pi)$ described above, one can realize single-qubit NOT gate,
i.e., $U_{NOT}=H_{z}(\pi)H_{inv}$. Certainly, this implemented
process needs to be adiabatic, otherwise the state will evolve along
one of the two Landau-Zener tunneling paths, which suppresses the
desirable complete population inversion between the two logic
states.

(ii) Implement the Hadamard gate. As another example, we set the
pulses sequence as (see Fig.~\ref{singlequbit}(b)): Stark pulse
precedes pump pulse to obtain the initial condition
$\vartheta(t_{0})=0$ and switches off prior to the pump pulse
resulting in $\vartheta(t_{f})=\pi/4$. This means that
\begin{eqnarray}
&&\!\!\!\!\!\!\!\!\!\!\!\!\!\!\!\!\!\!\!\!\!\! R_{h}:\left\{
\begin{split}
|\Psi(t_{0})\rangle=|0\rangle\xrightarrow{|\lambda_{-}(t)\rangle}
|\Psi(t_{f})\rangle=\frac{|0\rangle - |1\rangle}{\sqrt{2}}\\
|\Psi(t_{0})\rangle=|1\rangle\xrightarrow{|\lambda_{+}(t)\rangle}
|\Psi(t_{f})\rangle=\frac{|0\rangle + |1\rangle}{\sqrt{2}}
\end{split}
\right.\label{superposition passage}
\end{eqnarray}
This is the standard Hadamard gate operation.

\subsection{Physical implementation with single flux-biased phase qubit}

SJCs provide a favorable approach to implement QC due to its
nonlinearity. Especially, flux-biased phase qubits are typically
utilized to perform QC with superconducting
circuits~\cite{{Martinis},{Simmonds}}.

\begin{figure}[t]
\includegraphics[width=7cm,height=6cm]{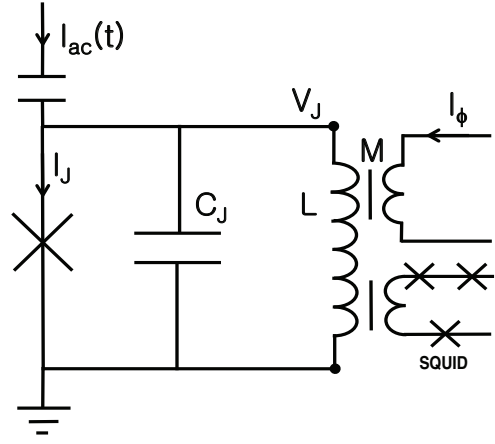}
\caption{Circuit schematic for a single flux-biased phase
qubit~\cite{Martinis}, where the $X$ symbol denotes the JJ.
$I_{ac}(t)$ is the microwave pump pulse and $I_{\phi}$ is the biased
dc-current.}\label{circuit}
\end{figure}

Typically, a flux-biased phase qubit is generated by a
superconducting loop (of inductance $L$) biased by a magnetic flux
$\Phi_e$ and interrupted by a Josephson junction (JJ) (with
capacitance $C$ and critical current $I_{0}$). The advantage of this
structure is that the generated qubit can be isolated well from the
strongly-dissipative bias leads~\cite{martinis3}. Using Kirchhoff's
law, the currents along all the branches of the circuit shown in
Fig.~\ref{circuit} satisfy the equation
\begin{equation}
I_{J}+I_{C}+I_{L}=I_{ac}(t).
\end{equation}
Using the Josephson current-phase relation $I_{J}=I_{0}\sin\delta$
and the voltage-phase relation $\dot{\delta}=2\pi V_{J}/\Phi_{0}$,
the above equation can be straightforward rewritten
as~\cite{ansmann2}
\begin{equation}
I_{0}\sin\delta(t)+C_{J}\frac{d}{dt}\frac{\Phi_{0}}{2\pi}\dot{\delta}+\int
\frac{V_{J}}{L}dt=I_{ac}(t).
\end{equation}
This equation can be further expressed by~\cite{Clarke}
\begin{equation}
C_{J}(\frac{\Phi_{0}}{2\pi})^2\ddot{\delta}+\frac{\partial U(\delta)}{%
\partial{\delta}}=0 ,\label{motioneq}
\end{equation}
with
\begin{eqnarray}
U(\delta)&=&-\frac{\Phi_{0}}{2\pi}I_{0}\cos\delta-\frac{\Phi_{0}}{2\pi}I_{ac}(t)\delta \nonumber\\
&&+\int\frac{\Phi_{0}}{2\pi
L}\left(\frac{\Phi_{0}}{2\pi}\delta-\Phi_{ex} \right)d\delta \nonumber\\
&=&E_{J}\left(\frac{(\delta-\phi_{b})^{2}}{2\lambda}-\cos\delta\right)-\frac{\Phi_{0}}{2\pi}I_{ac}(t)\delta.
\end{eqnarray}
Here, $\delta$ is gauge invariant phase difference (macroscopic
variable) of the JJ, $\Phi_{ex}=MI_{\phi}$ is the applied magnetic
flux, $E_{J}=I_{0}\Phi_{0}/2\pi$ the JJ coupling energy, and
$\lambda=2\pi I_{0}L/\Phi_{0}$, $\phi_{b}=2\pi\Phi_{ex}/\Phi_{0}$.
Also, $I_{\phi}=I_{\phi0}+I_{dc}(t)$ is the biased current with
$I_{\phi0}$ being the constant part and $I_{dc}(t)$ the
level-chirping part used for generating the Stark shift.

The above potential function $U(\delta)$ can be divided into two
parts: time-independent and time-dependent ones, i.e.,
\begin{equation}
U(\delta)= U_{0}(\delta)+V(t),
\end{equation}
with
\begin{eqnarray}
U_{0}(\delta)=E_{J}\left(\frac{(\delta-\phi_{b0})^{2}}{2\lambda}-\cos\delta\right),\nonumber\\
V(t)= -\frac{\Phi_{0} M}{2\pi L}I_{dc}(t)\delta
-\frac{\Phi_{0}}{2\pi}I_{ac}(t)\delta,\nonumber
\end{eqnarray}
where $\phi_{b0}=2\pi I_{\phi0}M/\Phi_{0}$.

Obviously, Eq.~(\ref{motioneq}), i.e., the equation for the gauge
invariant phase difference $\delta$, can be interpreted as that for
the motion of a particle with mass $m=C_{J}(\Phi_{0}/(2\pi))^2$
moving in the potential $U(\delta)$.
Certainly, the shape of this potential can be controlled by
adjusting the biased current $I_{\phi}$, which indirectly changes
magnetic flux though the loop. The bounded particle moving in the
potential would have discrete energy levels.
It is well known that all bound states of natural atoms/molecules
have definite parities, and therefore the so-called electric-dipole
selection rule determines all the possible transitions between the
selected levels. This rule forbids the transition between the states
with the same parity. However, in certain artificial atoms generated
by, e.g., the present SJCs, the bound states lose the definite
parities, and thus the electric-dipole transitions between arbitrary
two levels are possible. This provides a convenient way to design
the requirable pulses for implementing the above population
transfers.
\begin{figure*}[t]
\includegraphics[width=10cm,height=16cm,angle=270]{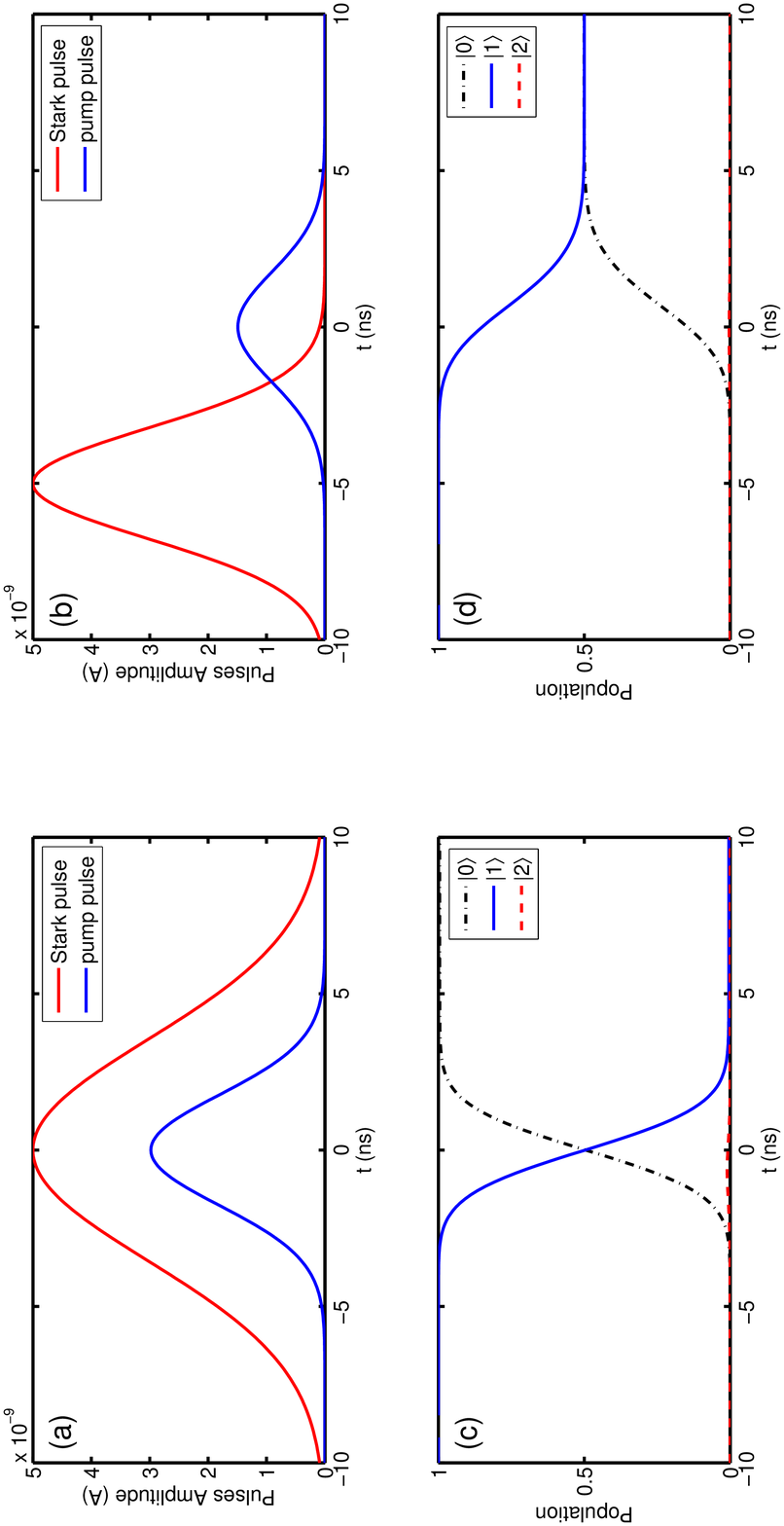}
\caption{(Color online) Population evolutions in a single
flux-biased phase qubit based on the proposed SCRAP. (left)
Population inversion operation achieved by using the adiabatic
pulses (a) with the parameters $\mathrm{I_{dc}(t)=5exp(-t^2/5^2)}$
nA, $\mathrm{\Omega(t)=2.98exp(-t^2/2.5^2)}$ nA. Here, the qubit
initially resides in the state $|1\rangle$, after the designed
SCRAP, $|0\rangle$ is fully populated. The population in the upper
level (i.e., $|2\rangle$) is significantly small (its largest value
is typically less than $1\%$), and thus during the SCRAP the
influence of this level is negligible. (right) Preparation of the
superposition state $(|0\rangle + |1\rangle)/\sqrt{2}$ from the
state $|1\rangle$ by the SCRAP with the adiabatic pulses
$\mathrm{I_{dc}(t)=5exp(-(t+5)^2/2.5^2)}$ nA,
$\mathrm{\Omega(t)=1.495exp(-t^2/2.5^2)}$ nA.}\label{singlequbit}
\end{figure*}

First, a proper magnetic flux is applied to let the junction has
several bound levels in the potential.
Usually, the lowest two levels with splitting-frequency
$\omega_{10}=(E_{1}-E_{0})/\hbar$ are selected to encode a JJ phase
qubit, and the third one $|2\rangle$ might be involved during the
qubit operations.

Second, in order to perform the expected SCRAP introduced above, a
microwave pump pulse $I_{ac}(t)=\varepsilon(t)\cos \,\omega_{10}t$
and a controllable Stark pulse $I_{dc}(t)$ are applied to couple the
qubit states, and chirp the qubit's transition frequency,
respectively. Under these drivings, the Hamiltonian of the above
flux-biased JJ reads~\cite{steffen2}
\begin{align}
\bar{H}_{s}(t)&=\bar{H}_{0}+V(t), \
\end{align}
with
\begin{align}
\bar{H}_{0}&=p^2/2m+U_{0}(\delta)=\sum_{i=0,1,2}E_{i}|i\rangle\langle
i|,\nonumber
\end{align}
and
\begin{align}
V(t)&=-(\Phi_{0}/2\pi)\Big[\frac{M}{L}I_{dc}(t)+I_{ac}(t)\Big]\delta&\nonumber \\
&=-(\Phi_{0}/2\pi)\Big[\frac{M}{L}I_{dc}(t)+I_{ac}(t)\Big]\sum_{i,j=0,1,2}|i\rangle
\langle j|\delta_{ij},&\nonumber
\end{align}
where $\delta_{ij}=\langle i|\delta|j\rangle$.
We assume that Stark shifts induced by the pump pulse are ignorable
compared to those induced by the Stark pulse, and also that
couplings (between the selected levels) induced by the Stark pulse
are negligible compared to those induced by the pump pulse. In the
interaction picture and under the usual rotating-wave appropriation,
the above Hamiltonian can be rewritten as
\begin{align}
\tilde{H}_{int}(t)&=\exp(it\bar{H}_{0})V(t)\exp(-it\bar{H}_{0})&  \nonumber \\
&=-\frac{\Phi_{0}}{2\pi}\left(
\begin{array}{ccc}
0 & \kappa\delta_{01} & 0 \\
\kappa\delta_{10} & \tilde{\Delta}_{1}(t) & \kappa\delta_{12}
\exp(i\nu t)
\\
0 & \kappa\delta_{21}\exp(-i\nu t) & \tilde{\Delta}_{2}(t) \\
\end{array}
\right),& \nonumber \\
\end{align}
where $\kappa=\Omega(t)/2,\nu=\omega_{10}-\omega_{21},\tilde{\Delta}%
_{1}(t)=MI_{dc}(t)(\delta_{11}-\delta_{00})/L, \tilde{\Delta}_{2}(t)=%
M I_{dc}(t)(\delta_{22}-\delta_{00})/L$.
For clarity, we return to the Schr\"odinger picture, and the above
equation can be rewritten as
\begin{align}
\bar{H}(t)&=-\frac{\Phi_{0}}{2\pi}\left(
\begin{array}{ccc}
0 & \kappa\delta_{01} & 0 \\
\kappa\delta_{10} & \tilde{\Delta}_{1}(t) & \kappa\delta_{12}
\\
0 & \kappa\delta_{21} & \tilde{\Delta}_{2}(t)+h\nu/\Phi_{0} \\
\end{array}
\right). &
\end{align}
With the experimental parameters~\cite{Simmonds,johnson2}
$I_{0}=8.351$ $\mu$A, $C_{J}=1.2$ pF, $L=168$ pH, $L/M=81$, and
$I_{\phi0}=923.7$ $\mu$A, one can numerically confirm that four
bound states (levels) exist in the left well of the potential
$U(\delta)$, and also $\delta_{00}=1.571$, $\delta_{11}=1.598$,
$\delta_{22}=1.633$, $\delta_{01}=\delta_{10}=0.076$,
$\delta_{12}=\delta_{21}=0.109$, $\delta_{02}=\delta_{20}=-0.006$,
$\omega_{10}/2\pi = 10.981$ GHz, $\omega_{21}/2\pi = 10.340$ GHz.

Fig.~\ref{singlequbit} shows the desirable single qubit operation.
(c) denotes that, under the pulse sequence in (a), the population in
the state $|1\rangle$ can be completely inverted to the state
$|0\rangle$. Inversely, if the state is initially prepared at the
state $|0\rangle$, then this pulse sequence will drive the system
evolving completely to the state $|1\rangle$. Clearly, during this
SCRAP for implementing the $\sigma_x$-rotation, leakage to
$|2\rangle$ (see the dashed red line in (c)) is really significantly
small and thus could be neglected. These numerical results also
confirm that the desirable population inversion can be finished
within the time interval $T\geq10$ ns, which is really rapid
compared to the sufficiently-long decoherence time (typically, e.g.,
$120$ ns~\cite{bialczak}).
Analogously, Fig.~\ref{singlequbit}(d) numerically confirms that the
desirable Hadamard gate operation can be demonstrated by applying
the pulse sequence in (b). Here, $|1\rangle$ is assumed to be
populated initially, then under the designed pulse sequence the
population is passed to the final state $(|0\rangle +
|1\rangle)/\sqrt{2}$ according to Eq.~(\ref{superposition passage}).

The SCRAP-based population transfer approach proposed here can be
directly utilized to implement the readout of the qubit with
significantly-high fidelity. Previously, the Josephson phase qubit
is read out by applying a readout pulse to fast lower the barrier of
the potential~\cite{cooper}. The aim of this operation is to quickly
enhance the tunneling probability of the upper level $|1\rangle$ for
being detected. Now, our population transfer approach provides
another way to read out the qubit. This can be achieved by
completely transferring the population of one of the logic states to
the readout state $|R\rangle$ with significantly-high tunneling
probability for detection. This approach is similar to
that~\cite{Martinis} by applying a $\pi$-pulse to resonantly drive
one of the logic states for evolving it to the readout state
$|R\rangle$. The difference in our scheme is that the duration of
the $\pi$-pulse is not required to be exactly designed, and also the
measurement fidelity could be significantly high. This is because
that the population of the selected logic state has been completely
transferred to the readout state $|R\rangle$ via the proposed SCRAP.

\section{Two-qubit gate operations in coupled Josephson phase qubits by SCRAP}

For the purpose of QIP, there must be lots of qubits coupled
together to form a quantum register. Fundamentally, any two-qubit
gate assisted with arbitrary single-qubit rotation generates an
universal set to produce any quantum computing circuit. In the
previous section, we have shown that the single-qubit
$\sigma_x$-rotation, phase-shift operation, and also the famous
Hadamard gate can be implemented in Josephson phase qubit by the
proposed SCRAP technique. In principle, any single-qubit rotation
can be generated by combining the typical single-qubit operations.
Now, in this section we show to implement a typical two-qubit gate,
i.e., iSWAP one, with two capacitively-coupled flux-biased Josephson
phase qubits~\cite{steffen}. Such a typical two-qubit gate has
already demonstrated by using the usual $\pi$-pulse
technique~\cite{bialczak}. The condition for exactly designing the
duration of the applied pulse will be relaxed in our scheme.

\begin{figure}[t]
\includegraphics[width=8.5cm,height=4.5cm]{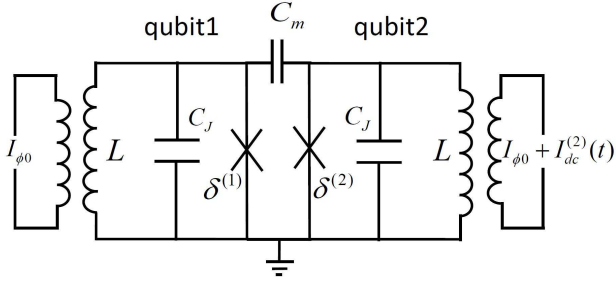}
\caption{Circuit diagram of the capacitively-coupled phase
qubits~\cite{johnson1}. Here, two flux-biased Josephson phase qubits
are coupled by the capacitance $C_m$. The time-dependent dc current
$I^{(2)}_{dc}(t)$ is applied to the second qubit to generate the
desirable Stark shift.}\label{circuit2}
\end{figure}

Without loss of generality, we consider a superconducting circuit
formed by two capacitively-coupled flux-biased JJs. Also, for
simplicity we assume that two junctions are identical and possess
the same energy structures (due to they are biased by the identical
magnetic fluxes). Here, the practically-existing capacitive coupling
could be served as the constant pump. An additional weak dc-current
(its amplitude is time-dependent) is applied to one of the junction
and serves as the required Stark pulse for chirping the levels of
the qubit (see Fig.~\ref{circuit2}). The Hamiltonian of this circuit
is~\cite{johnson1}
\begin{eqnarray}
\bar{H}_{12}(t)&=&\sum_{k=1,2}H_{0,k}+(2\pi /\Phi _{0})^{2}p_{1}p_{2}/\bar{C%
}_{m}\nonumber \\
&&-(\Phi _{0}/2\pi )\frac{M}{L}I_{dc}^{(2)}(t)\delta ^{(2)}  \nonumber \\
&=&\sum_{k=1,2}H_{0,k}+V_{1}+V_{2}.
\end{eqnarray}%
Here, $H_{0k}=(2\pi /\Phi
_{0})^{2}p_{k}^{2}/(2\bar{C}_{J})+E_{J}[(\delta^{(k)} -\phi
_{b})^{2}/(2\lambda )-\cos \delta^{(k)} ]$ describes the uncoupled
$k$th qubit with a renormalized junction capacitance
$\bar{C}_{J}=C_{J}(1+\zeta )$ with $\zeta =C_{m}/(C_{J}+C_{m})$.
Also, $C_{m}$ is the actual coupling capacitance between the two
qubits, $I_{dc}^{(2)}(t)$ is the chirping current applied to the
second junction. Furthermore, $V_{1}=(2\pi/\Phi
_{0})^{2}p_{1}p_{2}/\bar{C}_{m}$ represents the interaction between
two qubits with $\bar{C}_{m}=C_{J}(1+\zeta )/\zeta $ being the
effective coupling capacitance~\cite{kofman}, and $V_{2}=-(\Phi
_{0}/2\pi )MI_{dc}^{(2)}(t)\delta ^{(2)}/L$, related to the applied
chirping current, denotes the additional perturbation on the second
qubit.

Suppose that the chirping current is sufficiently weak, such that
the dynamics of each qubit is still safely limited within the
subspace $\emptyset _{k}=\{ |0_{k}\rangle, |1_{k}\rangle,
|2_{k}\rangle \}$, $\sum_{l=0}^{2}|l_{k}\rangle \langle l_{k}|=1$.
As a consequence, the circuit evolves within the total Hilbert space
$\emptyset_{k}=\emptyset_{1}\otimes \emptyset_{2}$.
Under the usual rotating-wave
appropriation, $V_{1}$ and $%
V_{2}$ can be rewritten as
\begin{align}
\tilde{V}_{1}(t) =&\exp (itH_{01})\exp (itH_{02})V_{1}\exp
(-itH_{01})\exp
(-itH_{02}) & \nonumber \\
=&(\frac{2\pi}{\Phi _{0}})^{2}\frac{1}{\bar{C}_{m}}\Big[\sum
|ij\rangle \langle ij|p_{ii}^{(1)}p_{jj}^{(2)}+\sum_{i\neq j}
|ij\rangle \langle
ji|p_{ij}^{(1)}p_{ji}^{(2)}&\nonumber \\
&+|02\rangle \langle 11|p_{01}^{(1)}p_{21}^{(2)}\exp (-i\omega
_{10}^{(1)}t)\exp (i\omega _{21}^{(2)}t)&  \nonumber \\
&+|11\rangle \langle 02|p_{10}^{(1)}p_{12}^{(2)}\exp (i\omega
_{10}^{(1)}t)\exp (-i\omega _{21}^{(2)}t)& \nonumber \\
&+|20\rangle \langle 11|p_{21}^{(1)}p_{01}^{(2)}\exp (i\omega
_{21}^{(1)}t)\exp (-i\omega _{10}^{(2)}t)&  \nonumber \\
&+|11\rangle \langle 20|p_{12}^{(1)}p_{10}^{(2)}\exp (-i\omega
_{21}^{(1)}t)\exp (i\omega _{10}^{(2)}t)\Big],&
\end{align}
and
\begin{eqnarray}
\tilde{V}_{2}(t)&=&\exp (itH_{02})V_{2}\exp (-itH_{02}) \nonumber\\
&=&-(\frac{\Phi _{0}}{2\pi })\frac{M}{L}I_{dc}^{(2)}(t) \sum
|ij\rangle \langle ij| \delta_{jj}^{(2)},
\end{eqnarray}
with $i, j=0, 1, 2$, and $p_{ij}=-i\hbar\langle
i|\frac{\partial}{\partial \delta}|j\rangle=-i\hbar p'_{ij}$.
\begin{figure*}[t]
\includegraphics[width=10cm,height=16cm,angle=270]{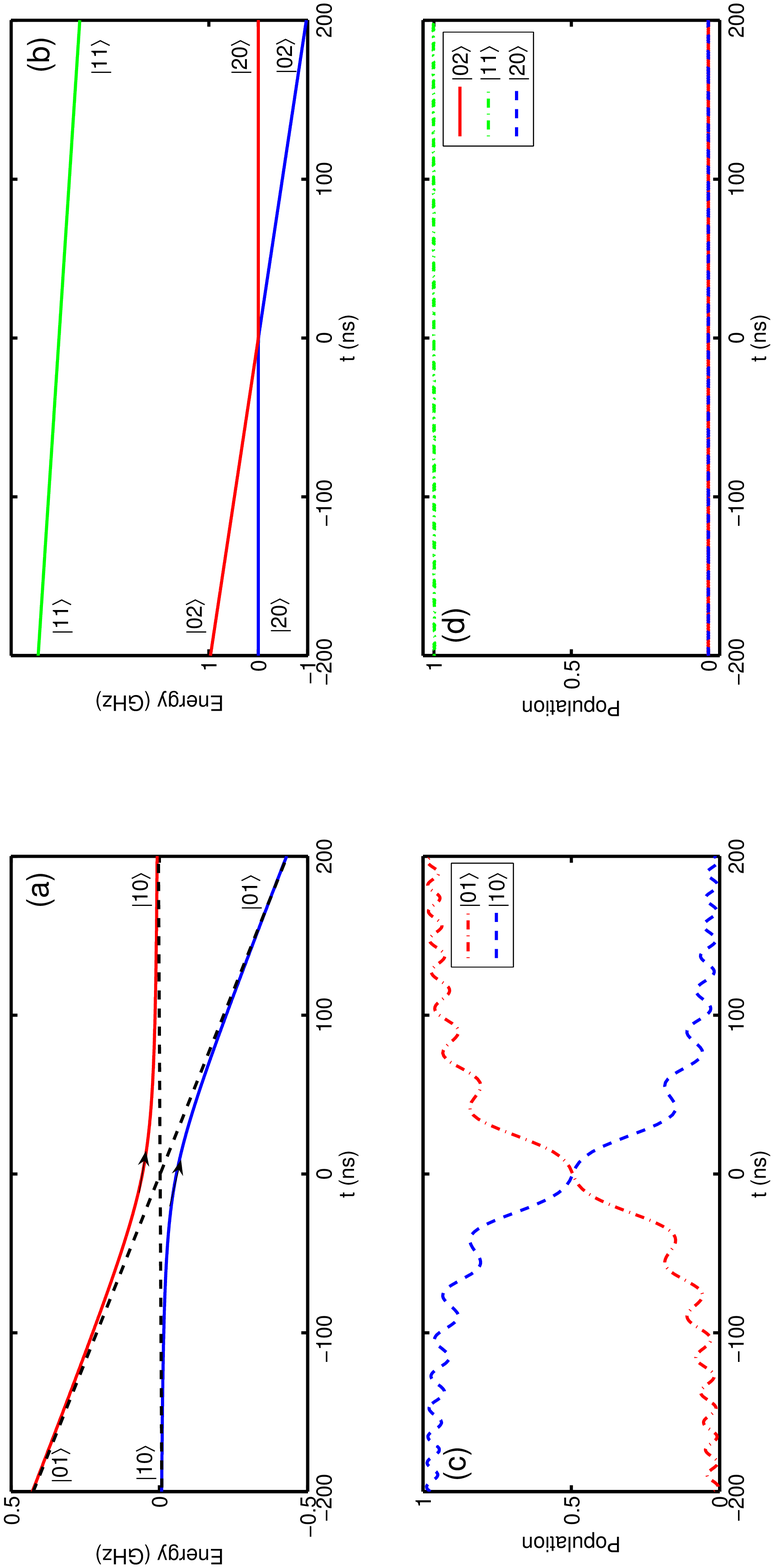}
\caption{(Color online) Population evolutions in two
capacitively-coupled qubits during the designed SCRAP: the Stark
pulse is generated by applying the weak dc-current $I_{dc}(t)=\gamma
t$ with $\gamma=2$ nA/ns. (a) Adiabatic passages and (c) population
swap between$|01\rangle$ and $|10\rangle$. (b) The adiabatic paths
and (d) the population evolution of the states in the subspace
$\Im_{3}=\{|02\rangle, |11\rangle, |20\rangle\}$ versus passage
time. Here, the system is initially prepared in the state
$|11\rangle$. One can see that during the passage the population in
this state is unchanged (dash-dotted green), and also the states
$|02\rangle$ (solid red) and $|20\rangle$ (dashed blue) are kept to
be unoccupied.}\label{twoqubit}
\end{figure*}
Under the condition that Stark shifts of the levels are
relatively-weak, the matrix elements of momenta
$p_{ij}^{(1)}=p_{ij}^{(2)}\equiv p_{ij}$. In fact, for the circuit
with effective coupling coefficient $\zeta=0.0017$ considered here,
our numerical calculations show that: $p'_{00}=0.271, p'_{11}=0.779,
p'_{22}=1.219, p'_{01}=p'_{10}=6.465, p'_{02}=p'_{20}=1.059,
p'_{12}=p'_{21}=8.761$.

In the interaction picture, one can easily check that the dynamics
of the system exists three invariant subspaces: (i) $\Im_{1}=\{
|00\rangle \}$; (ii) $\Im_{2}=\{ |01\rangle, |10\rangle \}$; and
(iii) $\Im_{3}=\{ |02\rangle, |11\rangle, |20\rangle \}$.
The first subspace $\Im_{1}$ includes only one quantum state
$|00\rangle$, the reduced Hamiltonian in this subspace reads
$\tilde{H}_{1}(t)=E_{00}(t)|00 \rangle \langle 00|$ with
\begin{equation}
E_{00}(t)=-\frac{\Phi_{0}}{2\pi}\frac{M}{L}I_{dc}^{(2)}(t)\delta_{00}^{(2)}+(\frac{2\pi
}{\Phi
_{0}})^{2}\frac{1}{\bar{C}_{m}}p_{00}^{(1)}p_{00}^{(2)}.\nonumber
\end{equation}
Certainly, if the system is initially prepared in the state
$|00\rangle$, then it always populates in this state. This implies
that we have the following evolution
\begin{equation}
|00\rangle\longrightarrow|00\rangle. \nonumber
\end{equation}
While, in the second subspace $\Im_{2}=\{a'=|01\rangle
,b'=|10\rangle \}$, the corresponding Hamiltonian can be written as
\begin{equation}
\tilde{H}_{2}(t)=\left(
\begin{array}{cc}
0 & \Omega _{a'b'} \\
\Omega _{b'a'} & \Delta _{b'b'}(t) \\
\end{array}%
\right),\label{swaph}
\end{equation}%
with
\begin{eqnarray}
\Omega _{a'b'} &=&\Omega _{b'a'} =(\frac{2\pi }{\Phi _{0}})^{2}\frac{1}{%
\bar{C}_{m}}p_{10}^{(1)}p_{10}^{(2)},\nonumber \\
\Delta _{b'b'}(t) &=&\frac{\Phi _{0}}{2\pi }\frac{M}{L}%
I_{dc}^{(2)}(t)(\delta _{11}^{(2)}-\delta _{00}^{(2)}).\nonumber
\end{eqnarray}
Returning to the Schr\"odinger picture, one can get the relevant
adiabatic paths and the population evolutions
(Fig.~\ref{twoqubit}(a, c)).
Of course, if the two qubits are exactly resonant, i.e., $\Delta
_{b^{\prime }b^{\prime }}(t)=0$, Hamiltonian Eq.~(\ref{swaph}) will
lead to a periodic swap of the populations between the states
$|01\rangle$ and $|10\rangle$~\cite{bialczak} with the period of
$\hbar \pi/(2\Omega_{a'b'})$. Certainly, the efficiencies of these
population transfers are sensitive to the evolution time $t$. In
order to overcome these evolution-time sensitivities for
implementing the desired swap operation, we apply an additional
dc-current $I_{dc}^{(2)}(t)=\gamma t$ with $\gamma=2$ nA/ns to chirp
the levels of the second qubit. If the system is initialized to be
$|10\rangle$, then after the adiabatic passage (blue line in
Fig.~\ref{twoqubit}(a)), $|01\rangle$ can be populated. This relaxes
the requirement of accurately designing the interaction time between
the qubits. Yes, such a passage needs a longer time but it still
could be finished within coherence time.
In the subspace $\Im_{3} =\{a=|02\rangle, b=|11\rangle,
c=|20\rangle\}$, the Hamiltonian expressed as
\begin{equation}
\tilde{H}_{3}(t)=\left(
\begin{array}{ccc}
E'_{0}(t) & \Omega_{ab}\exp(-i\theta t) & \Omega_{ac} \\
\Omega_{ba}\exp(i\theta t) & E'_{1}(t) & \Omega_{bc}\exp(i\theta t) \\
\Omega_{ca} & \Omega_{cb}\exp(-i\theta t) & E'_{2}(t) \\
\end{array}
\right).
\end{equation}
Here
\begin{eqnarray}
E'_{i}(t)&=&-(\frac{\Phi_{0}}{2\pi})\frac{M}{L}I_{dc}^{(2)}(t)
\delta_{jj}^{(2)}+ (\frac{2\pi}{\Phi_{0}})^2\frac{1}{\bar{C}_{m}}
p_{ii}^{(1)}p_{jj}^{(2)}, \nonumber
\end{eqnarray}
with $i\in \{0,1,2 \},j=2-i$,
and
\begin{eqnarray}
\Omega_{ab}&=&\Omega_{ba}=(\frac{2\pi}{\Phi_{0}})^2\frac{1}{\bar{C}_{m}}%
p_{10}^{(1)}p_{12}^{(2)},  \nonumber \\
\Omega_{ac}&=&\Omega_{ca}=(\frac{2\pi}{\Phi_{0}})^2\frac{1}{\bar{C}_{m}}%
p_{20}^{(1)}p_{02}^{(2)},  \nonumber \\
\Omega_{cb}&=&\Omega_{bc}=(\frac{2\pi}{\Phi_{0}})^2\frac{1}{\bar{C}_{m}}%
p_{12}^{(1)}p_{10}^{(2)},  \nonumber \\
\theta&=&\frac{E_{1}-E_{0}}{\hbar}-\frac{E_{2}-E_{1}}{\hbar}%
=\omega_{10}-\omega_{21}. \nonumber
\end{eqnarray}
Again, the corresponding Hamiltonian in the Schr\"odinger picture
could be written as
\begin{eqnarray}
\bar{H}_{3}(t)=\left(
\begin{array}{ccc}
E'_{0}(t)-\hbar\theta & \Omega_{ab} & \Omega_{ac} \\
\Omega_{ba} & E'_{1}(t) & \Omega_{bc} \\
\Omega_{ca} & \Omega_{cb} & E'_{2}(t)-\hbar\theta \\
\end{array}
\right).
\end{eqnarray}
The adiabatic paths and the corresponding population evolutions in
this subspace are also illustrated in Fig.~\ref{twoqubit}(b, d). As
what we can see that, if the population initially resides in the
state $|11\rangle$, then it always keeps in this state during the
adiabatic passage designed for implementing the inversion between
the $|10\rangle$ and $|01\rangle$. Since there is not any avoid
crossing between the state $|02\rangle$ (or $|20\rangle$) and the
state $|11\rangle$, the state $|02\rangle$ (or $|20\rangle$) should
not be populated during the above passage.

\section{Discussions and Conclusions}
In summary, we showed that populations could be adiabatically
transferred between the selected quantum states via SCRAP.
Typically, these transfers are insensitive to the details of the
applied pulses area. Based on these controllable population
transfers, fundamental quantum gates and typical superposition
states could be deterministically implemented. Our generic proposal
is demonstrated by the experimentally-existing flux-biased Josephson
phase qubits.

Our numerical results showed that, although the designed passages
are required to be adiabatic, the single-qubit population inversion
could still be finished within the duration as short as that of the
usual $\pi$-pulse. However, as shown in Fig.~(\ref{twoqubit}) that
the two-qubit logic gate operation with SCRAP technique takes
relatively long time~\cite{bialczak}, e.g., about $300$ns required
to finish the population inversion between $|01\rangle$ and
$|10\rangle$. Thus, longer decoherence is required for the coupled
superconducting circuits.
We also investigated the influence of other levels on the designed
adiabatic passages. It was shown that the population transfers are
really limited between the selected levels and the leakages to other
levels are negligible.

Note that the applied Stark pulse for chirping the levels should be
sufficiently weak, compared with the original bias for defining the
energy levels. For example, for realizing the above two-qubit
operation, the maximal change ratios of the relevant transition
frequencies (resulting from the applied Stark pulse) are estimated
as $(\omega_{10}^{+}-\omega_{10} )/\omega_{10}=-0.617\%$ and
$(\omega_{10}^{-}-\omega_{10} )/\omega_{10}=0.602\%$, respectively.
Here, $\omega_{10}^{\pm}$ denote the transition frequencies between
the states $|0\rangle$ and $|1\rangle$, due to the applying of a
Stark current (~$\pm400$ nA).
Also, the applied Stark pulses vary the transition matrix elements
very weak, such that the influences on the actions of the usual pump
pulses could be negligible. As a consequence, the pump pulses could
still resonantly interact with two selected levels.

\section*{Acknowledgments}
This work was supported in part by the National Science Foundation
grant No. 10874142, 90921010, and the National Fundamental Research
Program of China through Grant No. 2010CB92304.

\section*{APPENDIX}
\makeatletter
\renewcommand\theequation{A\@arabic\c@equation }
\makeatother \setcounter{equation}{0} Schr\"odinger equation in the
interaction picture reads
\begin{eqnarray}
i\frac{\partial|\Psi(t)\rangle}{\partial
t}=H'_{int}(t)|\Psi(t)\rangle.
\end{eqnarray}
In the new basis of $|\lambda_{-}(t)\rangle$ and
$|\lambda_{+}(t)\rangle$,
\begin{equation}
|\Psi(t)\rangle = v_{-}(t)|\lambda_{-}(t)\rangle +
v_{+}(t)|\lambda_{+}(t)\rangle .   \\
\end{equation}
This induces that,
\begin{eqnarray}
i\frac{\partial|\Psi(t)\rangle}{\partial t}
=v_{-}(t)H'_{int}(t)|\lambda_{-}(t)\rangle +
v_{+}(t)H'_{int}(t)|\lambda_{+}(t)\rangle \nonumber\\ \label{eqphi}
\end{eqnarray}
Multiplying the above equation left with $\langle\lambda_{-}(t)|$,
we get
\begin{eqnarray}
i\frac{dv_{-}(t)}{dt}&=&v_{-}(t)\mu_{-}(t)- i
v_{-}(t)\langle\lambda_{-}(t)|\frac{d}{dt}|\lambda_{-}(t)\rangle\nonumber \\
&&-i v_{+}(t)\langle\lambda_{-}(t)|\frac{d}{dt}%
|\lambda_{+}(t)\rangle \label{vnegative}
\end{eqnarray}
Similarly, multiplying Eq.~(\ref{eqphi}) left with
$\langle\lambda_{+}(t)|$, we have
\begin{eqnarray}
i\frac{dv_{+}(t)}{dt}&=&v_{+}(t)\mu_{+}(t)- i
v_{+}(t)\langle\lambda_{+}(t)|\frac{d}{dt}|\lambda_{+}(t)\rangle\nonumber \\
&&-i v_{-}(t)\langle\lambda_{+}(t)|\frac{d}{dt}%
|\lambda_{-}(t)\rangle \label{vpositive}
\end{eqnarray}
This implies that
\begin{eqnarray}
i\frac{d}{dt} \left(
\begin{array}{c}
v_{-}(t) \\
v_{+}(t) \\
\end{array}
\right) =M_{1} \left(
\begin{array}{c}
v_{-}(t) \\
v_{+}(t) \\
\end{array}
\right)\label{adeq},
\end{eqnarray}
with
\begin{align}
M_{1}=\quad\quad\quad\quad\quad\quad\quad\quad\quad\quad\quad\quad\quad\quad\quad\quad\quad\quad\quad\quad\quad
\quad\; \nonumber\\
\left(
\begin{array}{cc}
  \mu_{-}(t)-i\langle\lambda_{-}(t)|
  \frac{d}{dt}|\lambda_{-}(t)\rangle & -i\langle\lambda_{-}(t)|\frac{d}{dt}|\lambda_{+}(t)\rangle \nonumber \\
  -i\langle\lambda_{+}(t)|\frac{d}{dt}|\lambda_{-}(t)\rangle &
  \mu_{+}(t)-i\langle\lambda_{+}(t)|\frac{d}{dt}|\lambda_{+}(t)\rangle
\end{array}
\right).\nonumber
\end{align}
By using the expressions of $|\lambda_{+}(t) \rangle$ and $|
\lambda_{-}(t) \rangle$ in Eqs.~8(a,b), we have
\begin{subequations}
\begin{equation}
\langle\lambda_{-}(t)|\frac{d}{dt}|\lambda_{+}(t)\rangle=
\frac{d\vartheta(t)}{dt}
\end{equation}
\begin{equation}
\langle\lambda_{+}(t)|\frac{d}{dt}|\lambda_{-}(t)\rangle=-
\frac{d\vartheta(t)}{dt}
\end{equation}
\begin{equation}
\langle\lambda_{+}(t)|\frac{d}{dt}|\lambda_{+}(t)\rangle=0
\end{equation}
\begin{equation}
\langle\lambda_{-}(t)|\frac{d}{dt}|\lambda_{-}(t)\rangle=0
\end{equation}
\end{subequations}
So, $M_{1}$ can be simplified as
\begin{eqnarray}
M'_{1}= \left(
\begin{array}{cc}
  \mu_{-}(t) &
-i\frac{d\vartheta(t)}{dt}\\
i\frac{d\vartheta(t)}{dt} & \mu_{+}(t)
\end{array}
\right).
\end{eqnarray}
Suppose that the condition
\begin{equation}
|\dot{\vartheta}(t)|\ll \mu_{+}(t)-\mu_{-}(t) \label{adiabatic
condition}
\end{equation}
is satisfied, then $M'_{1}$ reduces to
\begin{eqnarray}
M''_{1}= \left(
\begin{array}{cc}
  \mu_{-}(t) & 0\\
   0 & \mu_{+}(t)
\end{array}
\right).
\end{eqnarray}
Thus, the solutions of Eq.~(\ref{adeq}) reads
\begin{eqnarray}
\left(
\begin{array}{c}
v_{-}(t) \\
v_{+}(t) \\
\end{array}
\right)=M_{2} \left(
\begin{array}{c}
v_{-}(0) \\
v_{+}(0) \\
\end{array}
\right),
\end{eqnarray}
with
\begin{eqnarray}
M_{2}= \left(
   \begin{array}{cc}
       \exp(-i\int_{0}^{t}\mu_{-}(t')dt') & 0 \\
       0 & \exp(-i\int_{0}^{t}\mu_{+}(t')dt') \\
   \end{array}
\right).\nonumber
\end{eqnarray}
Returning to the new (adiabatic) basis, the above adiabatic
evolution solution takes the form
\begin{eqnarray}
|\Psi(t)\rangle &= v_{-}(0)\exp(-i\int_{0}^{t}\mu_{-}(t')dt')
|\lambda_{-}(t)\rangle+&\nonumber\\
&v_{+}(0)\exp(-i\int_{0}^{t}\mu_{+}(t')dt')|\lambda_{+}(t)\rangle.&
\end{eqnarray}
Here, the additional phases in adiabatic states can be regarded as
the dynamical phases produced during the adiabatic passages.

Next, we show what an adiabatic condition should be satisfied by
controlling Stark and pump pulses.

First, from the definition below Eq.~(\ref{Hintpp}),
\begin{eqnarray}
\qquad\qquad\tan 2\vartheta(t)&=&\frac{\Omega(t)}{\Delta(t)}=
\frac{2\tan \vartheta(t)}{1-\tan^{2}\vartheta(t)},
\end{eqnarray}
we have
\begin{eqnarray}
\frac{d}{dt}\tan 2\vartheta(t)&=&2
(1+\tan^{2} 2\vartheta(t))\frac{d\vartheta}{dt}\nonumber\\
&=&\frac{\frac{d\Omega(t)}{dt}\Delta(t)-\Omega(t)\frac{d\Delta(t)}
{dt}}{\Delta^{2}(t)},
\end{eqnarray}
and
\begin{eqnarray}
\frac{d\vartheta(t)}{dt}&=&
\frac{\frac{d\Omega(t)}{dt}\Delta(t)-\Omega(t)\frac{d\Delta(t)}{dt}}
{2\Delta^{2}(t)(1+\tan^{2} 2\vartheta(t))}\nonumber\\
&=&\frac{\frac{d\Omega(t)}{dt}\Delta(t)-\Omega(t)\frac{d\Delta(t)}{dt}}
{2(\Delta^{2}(t)+\Omega^{2}(t))}.
\end{eqnarray}
The adiabatic condition Eq.~(\ref{adiabatic condition}) implies that
\begin{align}
\frac{1}{2}\left|\Omega(t)\frac{d\Delta(t)}{dt}-\Delta(t)
\frac{d\Omega(t)}{dt}\right|\ll\left(\Delta^{2}(t)+\Omega^{2}(t)\right)^{3/2}.
\end{align}
Therefore, smooth pulses, long interaction time, and large Rabi
frequency and detuning are needed to satisfy the desirable adiabatic
condition.

\end{document}